# Astrophysics Conducted by the Lunar University Network for Astrophysics (LUNAR) and the Center for Lunar Origins and Evolution (CLOE)

*Science Of, On, and From the Moon undertaken by the NASA Lunar Science Institute*

### LUNAR

*Director: Dr. Jack Burns (University of Colorado, Boulder)*
*Deputy Directory: Dr. Joseph Lazio (Jet Propulsion Laboratory/California Institute of Technology)*

### CLOE

*Director: Dr. William Bottke (Southwest Research Institute)*

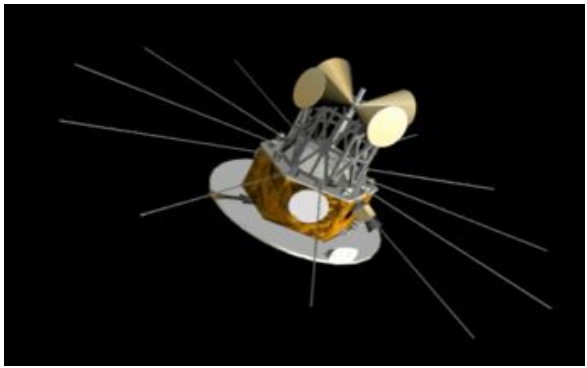
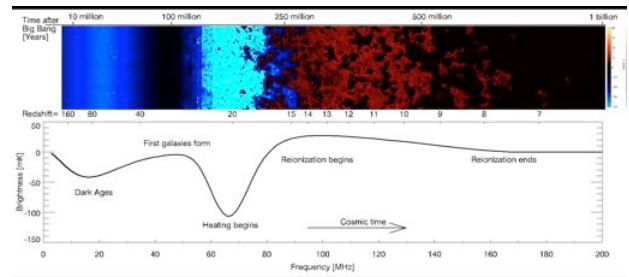
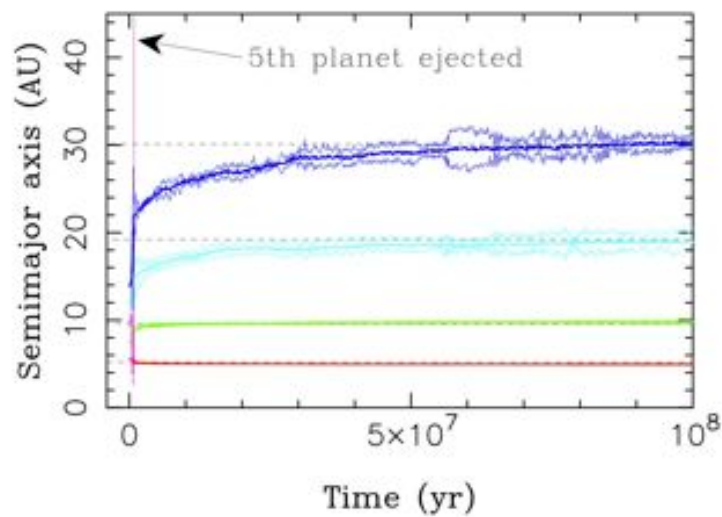

# Executive Summary

The Moon is a unique platform from and on which to conduct astrophysical measurements. The Lunar University Network for Astrophysics Research (LUNAR) and the Center for Lunar Origins and Evolution (CLOE) teams within the NASA Lunar Science Institute (NLSI) are illustrating how the Moon can be used as a platform to advance important goals in astrophysics. Of relevance to Astrophysics and aligned with NASA strategic goals, all three of the primary research themes articulated by *New Worlds, New Horizons in Astronomy & Astrophysics* are being addressed by LUNAR and CLOE:

**Probing Cosmic Dawn:** What were the first objects to light up the Universe? Determining the nature of and epoch of formation for the first luminous sources will be the focus of multi-wavelength observations for this decade, and potentially beyond. The evolution of the intergalactic medium (IGM), in response to the first luminous objects, can be tracked via the hyperfine spin-flip (21 cm) transition of neutral hydrogen. There is potentially even a signal *prior* to the formation of the first stars due to the different cooling rates of the IGM and the cosmic microwave background.

The Universal expansion redshifts the 21 cm line to much longer wavelengths, to a portion of the radio spectrum used extensively by both civil and military transmitters and for which ionospheric absorption becomes important. The far side of the Moon is shielded from terrestrial emissions and represents a unique location in the inner solar system for observing the redshifted 21 cm line during and before Cosmic Dawn. Members of the LUNAR team have been instrumental in developing the theoretical foundation for these observations. Concurrently, the LUNAR team has been pursuing a concept for a lunar orbiting spacecraft, the Dark Ages Radio Explorer, and technology development for future lunar surface telescopes.

**Understanding New Worlds:** "Can we find another planet like Earth orbiting a nearby star?" Searching for and understanding how planetary systems form and the processes shaping them will be the focus of multi-wavelength observations for this decade, and potentially beyond. Already, thousands of planets and planetary candidates are known. Characterizing these planets, as well as understanding what these planetary systems tell us about the formation and evolution of our solar system is of increasingly importance.

The CLOE team is combining the Moon's cratering record with simulations of the early solar system to understand its early evolution. Strikingly, the CLOE team is finding that the Late Heavy Bombardment of the inner solar system is most easily explained only if there was considerable dynamic evolution, including the likely ejection of a fifth planet. In addition, members of the LUNAR team are exploring how nearby planets might be both detected and characterized by means of magnetically generated emissions.

**Physics of the Universe**: What are the limits of physical laws? Astronomy offers the opportunity of testing physical laws in a way that cannot be accomplished in terrestrial laboratories. The LUNAR team has been using lunar laser ranging to make high precision measurements of the Earth-Moon distance (to 1 part in $10^{12}$) using retroreflectors emplaced on the Moon's surface during the *Apollo* era. The 40 year record of distance measurements now provides for some of the highest constraints on general relativity and other theories of gravity.

# 1) Probing Cosmic Dawn

**Project Leaders:** Dr. Joseph Lazio (Jet Propulsion Laboratory/California Institute of Technology)
Dr. Steven Furlanetto (University of California, Los Angeles)
Dr. Jack Burns (University of Colorado, Boulder)

## a) Recommendations from *New Worlds, New Horizons in Astronomy and Astrophysics*

The *New Worlds, New Horizons in Astronomy & Astrophysics* Decadal Survey (NWNH) identified "Cosmic Dawn" as one of the three science objectives guiding the science program for the decade between 2010 and 2020. In the science program articulated in NWNH (Chapter 2), the **Epoch of Reionization** (EoR) was identified as a science frontier discovery area that could provide the opportunity for "transformational comprehension, i.e., discovery," and **"What were the first objects to light up the Universe and when did they do it?"** was identified as a science frontier question in the Origins theme.

While our primary focus is on the current astronomy Decadal Survey, using the Moon as a platform for probing Cosmic Dawn via low radio frequency astronomy observations has been recognized in other NRC reports and community documents. As recent examples, both the NRC report *The Scientific Context for the Exploration of the Moon* and "The Lunar Exploration Roadmap: Exploring the Moon in the 21st Century: Themes, Goals, Objectives, Investigations, and Priorities (v. 1.1)" produced by the Lunar Exploration Analysis Group (LEAG) discuss the scientific value of a lunar radio telescope. Further, and importantly, as also discussed by *The Scientific Context for the Exploration of the Moon*, the "scientific rationale for lunar science and its goals and recommendations are independent of any particular programmatic implementation."

## b) Observations of the Highly Redshifted 21 cm Hydrogen Line

Following recombination at a redshift of $z \approx 1100$, the Universe entered a largely neutral state in which the dominant baryonic component of the intergalactic medium (IGM) is neutral hydrogen (H I). The approach of the LUNAR team is to use the (redshifted) spin-flip (21 cm) hyperfine transition of H I to track the influence of the first ionizing and heating sources on the IGM.

By a redshift of $z \sim 7$, observations with a combination of ground- and space-based telescopes are showing that the precursors of modern-day galaxies exist and are beginning to re-ionize their surroundings; analysis of WMAP data suggests that the EoR is likely to be an extended process, potentially beginning at $z \sim 12$. Future observations with *JWST* and the Atacama Large Millimeter/Submillimeter Array (ALMA) are likely to be able to probe ionizing sources to $z \approx 15$–20. As Figure 1 illustrates, the 21 H I line produces a signal potentially over the redshift range of approximately 6–100, from when the excitation (spin) temperature of the H I transition tracks that of the gas kinetic temperature ($z \sim 100$) to the late times of the EoR ($z \sim 6$). In between this range, the strength of the H I signal depends upon the temperature of the gas and the fraction of neutral hydrogen, quantities that both are impacted by the time and rate at which the first ionizing and heating sources form.

### c) The Moon as a Platform for 21 cm Cosmology

The lunar *farside* is potentially the only site in the inner solar system for high-precision 21 cm cosmology, as we discuss below. A number of ground-based projects are exploring detection of the 21 cm signal from various redshift epochs. Ultimately, however, we expect that evolution of the field of 21 cm cosmology will follow that of the successful history of measurements of the cosmic microwave background (CMB), namely, CMB studies can be conducted from the ground or from balloons, but the space environment substantially improves the scientific return. Terrestrial CMB instruments including BOOMERANG, DASI, CBI, and others developed fundamental techniques and illuminated key scientific uncertainties, but the COBE, WMAP, and Planck missions have capitalized on the space environment to provide definitive measurements.

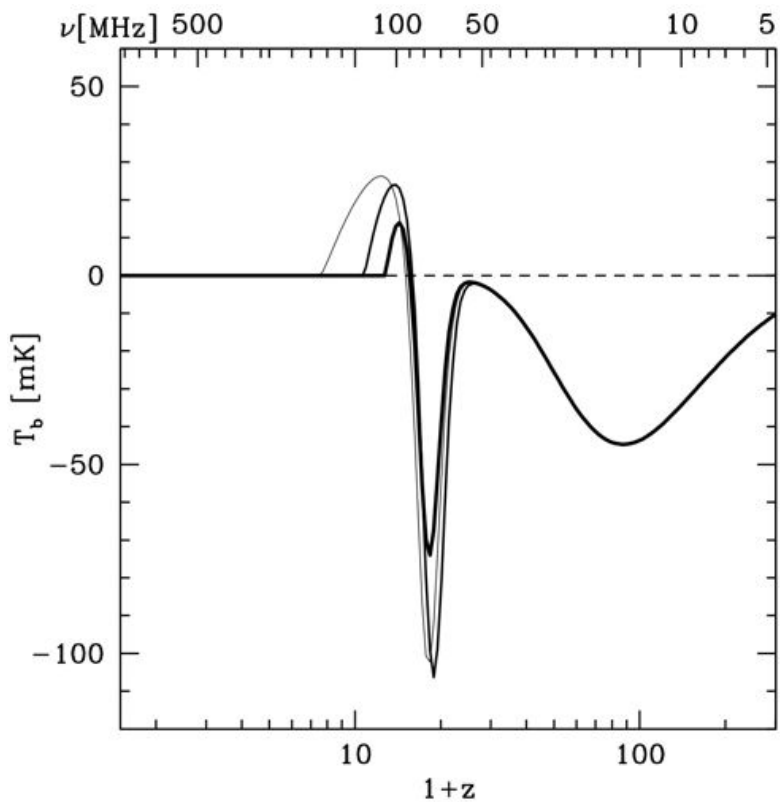

Figure 1. Evolution of the 21 cm global signal, relative to the temperature of the cosmic microwave background (CMB). The ordinate shows the brightness temperature of the (redshifted) 21 cm signal, and the bottom (top) abscissa shows the redshift (frequency, given the rest frame frequency of 1420 MHz). The thick black curve shows a notional model, with the first star formation occurring at $z \sim 30$, and in which the IGM is nearly completely ionized by a redshift of 6. Other curves illustrate a range of models for galaxy formation consistent with the WMAP constraints on the electron opacity. (From Pritchard & Loeb 2008) Three general epochs can be defined, the Dark Ages ($z \sim 30–200$), the First Stars and Cosmic Dawn ($z \sim 15–30$), and the Epoch of Reionization ($z \sim 6–15$).

#### i) No Human-generated Interference

The majority of the integrated radio power emitted by our civilization is at the relevant frequency range (< 150 MHz) for 21 cm cosmology (Loeb & Zaldarriaga 2007). The FM radio band is at 88–107 MHz, and Digital TV channels and myriad other signals, both civil and military, also exist in the relevant frequency range. At the Murchison Radio Observatory in Western Australia, a government-protected site, several FM radio signals are persistently detected at the 1 K level (1000 times the limit needed to detect the 21 cm cosmology signal) due to reflections from meteors and aircraft (Bowman et al. 2008; Rogers & Bowman 2008; Bowman & Rogers 2010a,b). Further, because of ionospheric refraction, interference in the HF band (< 30 MHz) used for international communication is essentially independent of location on Earth. Terrestrial transmitters can be orders of magnitude ($\sim 10^{12}$) stronger than the H I signals and are detectable at some level even at remote locations on Earth (Figure 2).

The Radio Astronomy Explorer-2 (RAE-2) and the *Apollo* Command Modules had radio systems that bracketed the relevant frequencies. Both observed a complete cessation of terrestrial radio emission while in the radio quiet zone above the lunar farside (Alexander et al. 1976). Moreover, reflections of terrestrial interference from other spacecraft in view of the lunar farside (e.g., at the Sun-Earth L2 point) will be at a negligible level (< 1 nK).

### i) No (Permanent) Ionosphere

The Earth's ionosphere is sufficiently dense to add two complications to ground-based 21 cm cosmology efforts. First, any plasma displays a frequency-dependent absorption of increasing magnitude, down to a critical frequency below which the plasma transmits no electromagnetic radiation. In the case of the Earth's ionosphere, the critical frequency is typically around 10 MHz, but Rogers (2011) has shown that the amount of absorption is both variable and comparable to the magnitude of the expected 21 cm cosmological signal as high as 100 MHz.

Second, density variations within the terrestrial ionosphere induce phase errors that limit radio observations (in addition to simply reflecting interference from distant transmitters, Figure 2). These phase errors form a significant fraction of the error budget in the recent 74 MHz Very Large Array (VLA) Low frequency Sky Survey (VLSS, Cohen et al. 2007), even after the development of new algorithms for ionospheric mitigation.

While the Moon has a plasma layer due to solar irradiation during the lunar day, its density, and therefore its critical frequency, is much lower than the Earth's. The critical frequency is not expected to exceed 1 MHz, and typically is closer to 0.3 MHz. Further, this ionized layer disappears during lunar night.

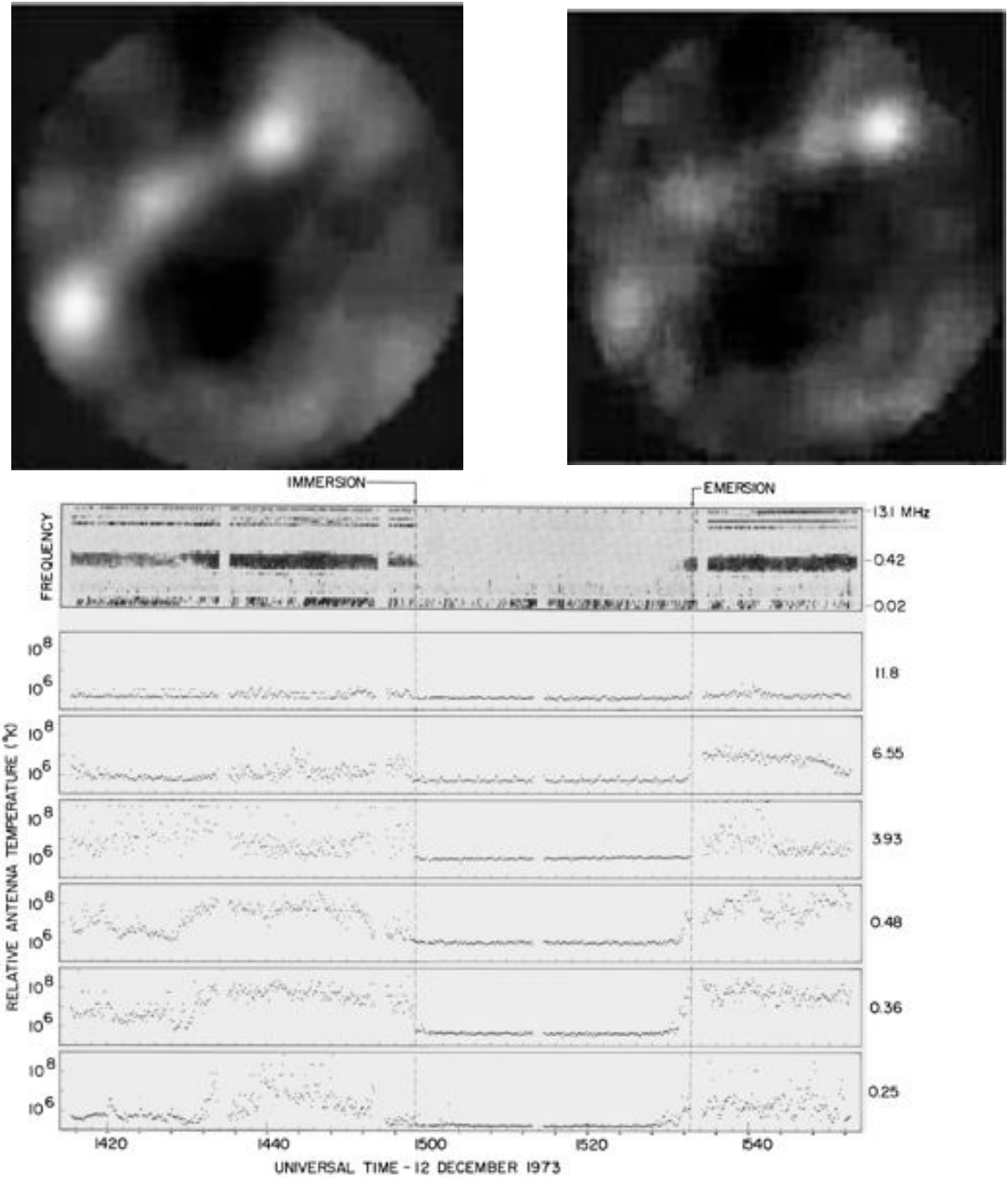

Figure 2. The highest sensitivity 21 cm cosmology observations will require shielding from terrestrial interference, such as that shown here. (*Top Left*) Radio interference enabled by the Earth's atmosphere. An all-sky, 60 MHz image acquired by the Long Wavelength Demonstrator Array in New Mexico. The Galactic plane slopes diagonally from the upper right to the lower left and the sources Cyg A and Cas A are visible as is a general enhancement toward the inner Galaxy. (*Top Right*) An image acquired seconds later. The dominant source (upper right) is a reflection of a TV station hundreds of kilometers away from an ionized meteor trail in the upper atmosphere. (*Bottom*) Lunar occultation of the Earth as observed by RAE-2 (Alexander et al. 1975). The top panel is a dynamic spectrum showing intensity as a function of time and frequency. The remaining panels show the intensity at a specific frequency as a function of time. Clearly apparent is the drop in intensity when RAE-2 was behind the Moon.

### ii) Shielding from Solar Radio Emission

The Sun's proximity makes it the strongest celestial source at these frequencies when it is bursting (Figure 3). Solar radio bursts are many orders of magnitude stronger than the expected H I signals. Within the solar system, the only mitigation for solar radio emissions is physical shielding. A free-flying mission would require a metallic screen of substantial size to shield it from solar radio emissions, including diffraction around the edges of the screen. Such a shield, while possible, would present a significant mass penalty to any mission and a risk from the perspective of deployment. Such shielding is readily accomplished by observing during lunar night and, while the same is true for the surface of the Earth, interference and ionospheric effects continue to occur during terrestrial night.

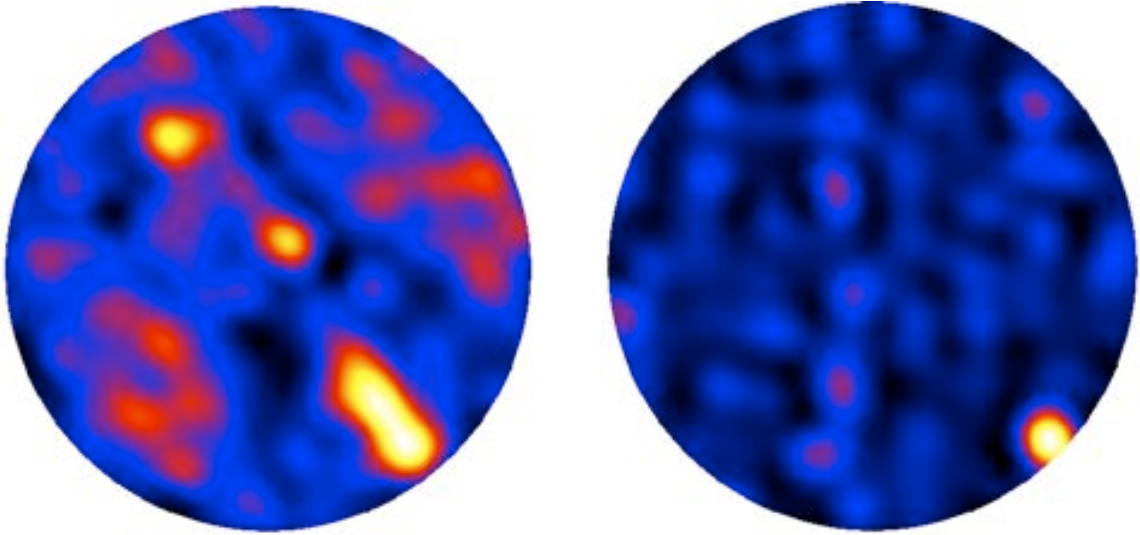

**Figure 3. The Sun as a radio source.** (*Left*) An all-sky image from the Long Wavelength Demonstrator Array in New Mexico at 74 MHz. In this image, the Galactic plane slopes diagonally from the upper left to the lower right, the inner Galaxy is to the lower right, Cyg A is in the center of the image, and Cas A is toward the upper left. This image was acquired in December, when the Sun was in the constellation Sagittarius (toward the inner Galaxy). (*Right*) The sky later the same day during a solar flare. Due to the finite dynamic range available, the only source visible is the Sun.

### d) Theoretical Developments by the LUNAR Team

The 21 cm brightness temperature of an IGM gas parcel at a redshift $z$, relative to the CMB, is (Madau et al. 1997; Furlanetto et al. 2006)

$$\delta T_b \approx 25 \text{ mK } x_{HI} (1 + \delta) [(1 + z)/10]^{1/2} [1 - T_{CMB}(z)/T_s] [H(z)/(1 + z)/dv_{||}/dr_{||}]$$

where $x_{HI}$ is the neutral fraction, $\delta$ is the fractional IGM overdensity in units of the mean, $T_{CMB}$ is the CMB temperature, $T_s$ is the spin (or excitation) temperature of this transition, $H(z)$ is the Hubble constant, and $dv_{||}/dr_{||}$ is the line-of-sight velocity gradient.

All four of these factors contain unique astrophysical information. The dependence on $\delta$ traces the development of the cosmic web, while the velocity factor incorporates line-of-sight "redshift-space distortions" that separate aspects of the cosmological and astrophysical signals. The other two factors depend strongly on the ambient radiation fields in the early Universe: the ionizing background for $x_{HI}$ and a combination of the ultraviolet background (which mixes the 21 cm level populations through the Wouthuysen-Field effect) and the X-ray background (which heats the gas) for $T_s$.

Figure 1 shows a model for the sky-averaged 21-cm spectrum during the Dark Ages, measured relative to the CMB (Furlanetto et al. 2006). Several epochs can be identified, but these are currently essentially unconstrained by observations. One of the goals of the LUNAR team is to develop the technologies for making such observations.

**The Dark Ages ($z > 30$)**: Before the first stars and galaxies formed, the hydrogen gas was influenced only by gas collisions and absorption of CMB photons. The gas cooled rapidly as the Universe expanded, and the resulting cold temperatures cause the 21 cm signal to appear in absorption. However, as the Universe expanded the decreased gas density reduced the collision rate, and absorption of CMB photons drove the spin temperature into equilibrium with the CMB. This caused the 21 cm absorption signal to decrease and eventually disappear by $z \sim 30$.

**First Stars and Cosmic Dawn ($30 > z > 22$)**: Shortly thereafter, the first stars appeared ($z \sim 30$). Their radiation "turned on" the 21 cm signal by triggering absorption and re-emission of ultraviolet photons through the Wouthuysen-Field mechanism. This drove the spin temperature toward the cold temperatures characteristic of IGM gas, sending the 21 cm signal into a deep absorption trough.

**First Accreting Black Holes ($22 > z > 13$)**: Black holes likely formed at this time, e.g., as remnants from the first stars. Due to their intense gravity, gas falling into them would be accelerated, shock heated, and begin to radiate X-rays. At this time, the IGM gas was extremely cold ($\sim 10$ K) and the energetic X-ray photons from the accreting black holes began to heat it. This heating transformed the spin-flip signal from absorption into emission as the gas became hotter than the CMB.

**Hot, Bubble Dominated Epoch ($13 > z > 6$)**: Also known as the **Epoch of Reionization**, once the gas became hot, the emission saturated, until photons from these stars and black holes started ionizing the gas in giant bubbles within the IGM. The rapid destruction of the neutral gas during reionization then eroded the 21 cm signal.

**Fully Ionized Universe ($z < 6$)**: Once the Epoch of Reionization completed and the neutral gas was destroyed, the 21cm spin-flip signal disappeared.

In what follows, we describe the two different approaches that have been developed, summarizing material that is reviewed in more depth by LUNAR team members (Furlanetto et al. 2006; Pritchard & Loeb 2008; Pritchard & Loeb 2011).

### i) The Sky-Averaged 21 cm Signal

The sky-averaged 21 cm spectrum (Figure 1) is the most basic quantity of physical interest. Thanks to the cosmological redshift, each observed frequency corresponds to a different cosmic epoch, so this spectrum allows us to trace the development of structure in the Universe. Figure 1 shows that the different eras identified in the previous section imprint distinct features—especially turning points—on this background spectrum. Crucially, our current understanding of the properties of the first galaxies and black holes is especially poor, with reasonable models providing variations over several orders of magnitude in their parameters. Pritchard & Loeb (2010) built upon earlier analytic models by Furlanetto (2006) to quantify this uncertainty and determine the relevant range of radio frequencies where these features might occur.

The primary challenge in observing this sky-averaged background is separating the cosmological signal from the low-frequency radio background. For many years, this challenge was thought to be insurmountable, but recent advances in instrumentation and analysis have offered a great deal of hope. Harker et al. (2012) developed a detailed model of global signal observations accounting for instrumental effects, foregrounds, and the 21 cm signal and by applying a Markov-Chain Monte-Carlo (MCMC) algorithm explored the achievable bounds on the positions of the signal turning points. This modeling will form the basis of the DARE data analysis pipeline and is in the process of being tested against data from EDGES (Bowman & Rogers 2010). Ultimately, observations of the global 21 cm signal could pin down the times when key transitions in the history of the Universe occurred, e.g., the heating of the IGM from cold to hot. Members of the LUNAR team are leading the way in developing both theoretical and observational tools in this field.

### ii) Fluctuations in the 21 cm Signal and Power Spectral Measurements

Complementing sky-averaged experiments are radio interferometric observations of 21 cm fluctuations. This approach allows one to measure the properties of individual structures, such as the ionized bubbles that fill the Universe during reionization, either by mapping them or through their statistical properties. It is therefore much more powerful than the sky-averaged signal, but it is also much more difficult to measure because the signal from each structure is extraordinarily small. Experimental data is just starting to become available from the EDGES, PAPER, and the GMRT, which will guide understanding of instrumental systematics in the future.

Members of LUNAR have been active in developing the theoretical tools needed to predict the 21 cm fluctuations. As noted earlier, these are sourced by both density and radiation emitted from galaxies (Madau et al. 1997). Analytic models show that the mean amplitudes of these fluctuations have several peaks through cosmic time, roughly corresponding to the same epochs outlined earlier. However, the spatial properties of these fluctuations also allow one to learn about the detailed properties of the radiating sources, including their masses, luminosities, and clustering properties.

From a theoretical standpoint, the primary challenges to understanding these signals are our poor knowledge of the properties of high-redshift galaxies and the huge dynamic range required to study both these sources (which are just a few kpc across) and the radiation fields that determine the spin-flip signal (which can vary across hundreds of Mpc). Numerical simulations of the representative volumes ($\sim$ Gpc$^3$) are currently numerically expensive and have yet to reach the required dynamic range (except for those that focus solely on reionization). Mesinger & Furlanetto (2007) have pioneered a different semi-numerical approach that makes use of analytic approximations for reionization and the radiative transfer of X-rays and Lyman series photons to realize 21 cm brightness fluctuations in almost arbitrarily large volumes. The analytic approximations have been validated against reionization simulations and work is in progress to test them in the early stages of the cosmic dawn. The code for these calculations—`21cmFast` (Mesinger et al.

2011)— has been made publicly available and is being used by a number of experimental groups around the world. The speed with which semi-numerical codes can produce detailed realizations of the 21 cm signal makes it a good fit for statistical data analysis, which must compare data to many different theoretical models.

Finally, the recent implementation of radiative transfer algorithms in numerous cosmological hydrodynamics codes has led to a dramatic improvement in studies of feedback in various astrophysical environments. However, because of methodological limitations and computational expense, the spectra of radiation sources are usually sampled at only a few, generally evenly spaced, discrete emission frequencies. Using 1D radiative transfer calculations, Mirocha et al. (2012) investigated the discrepancies in gas properties surrounding model stars and accreting black holes that arise solely due to spectral discretization. They found that even in the idealized case of a static, uniform density field, commonly used discretization schemes overestimate hydrogen ionization by factors of approximately 2 and underestimate temperatures by up to an order of magnitude at large radii. The consequences are most severe for radiative feedback operating on large (~ Mpc) scales, dense clumps of gas, and media consisting of multiple chemical species. They have developed a method for optimally constructing discrete spectra and show that for two test cases of interest, carefully chosen 4-bin spectra can eliminate errors associated with frequency resolution to high precision. Applying these findings to fully 3D radiation-hydrodynamic simulations of the early Universe, they found substantially altered H II region sizes and morphologies around first stars and black holes, and therefore a sizable impact on their associated observable signatures.

These signals will constrain models of the first luminous sources to form in the Universe, and theoretical progress on predicting them has already informed work on the high-redshift Universe in other contexts. For example, the same ultraviolet background that "turns on" the 21 cm background also suppresses primordial star formation throughout the Universe, so observing the background helps us to understand the earliest phases of galaxy formation (Holzbauer & Furlanetto 2012). Moreover, the recently recognized velocity offset between gas and dark matter in the early Universe substantially delays structure formation, and it may dramatically amplify the 21 cm background itself (Dalal, Pen, & Seljak 2010). As our understanding of the high-redshift Universe improves, we will be able to sharpen predictions for these signals and hence focus the experimental effort more fully on the relevant observational questions.

## e) Observational Concept Developments by the LUNAR Team

### i) Dark Ages Radio Explorer (DARE)

The Dark Ages Radio Explorer (DARE) is a concept for a lunar-orbiting, cosmology mission (Figure 4, Burns et al. 2012). DARE's science objectives [§0] include (1) When did the first stars form? (2) When did the first accreting black holes form? (3) When did Reionization begin? (4) What surprises does the end of the Dark Ages hold (e.g., Dark Matter decay)? DARE will use the highly-redshifted hyperfine 21 cm transition from neutral hydrogen to track the formation of the first luminous objects by their impact on the intergalactic medium during the end of the Dark Ages and during Cosmic Dawn (redshifts $z$ = 11–35, corresponding to a radio bandpass of 40–120 MHz). By measuring the sky-averaged spin temperature of neutral hydrogen at the unexplored epoch 80–420 million years after the Big Bang, DARE will provide the first evidence of the earliest stars and galaxies to illuminate the cosmos and testing our models of galaxy formation.

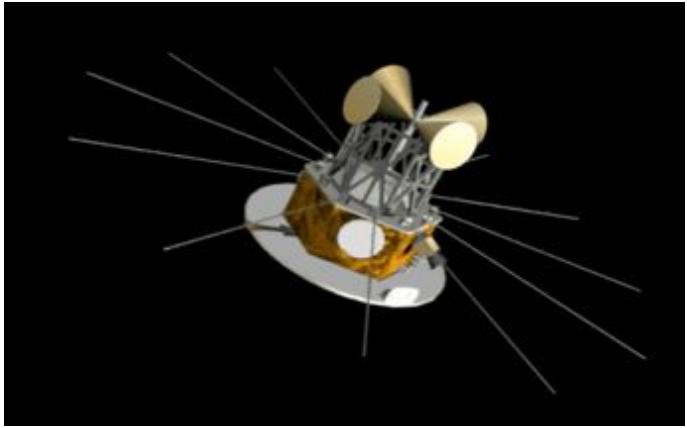
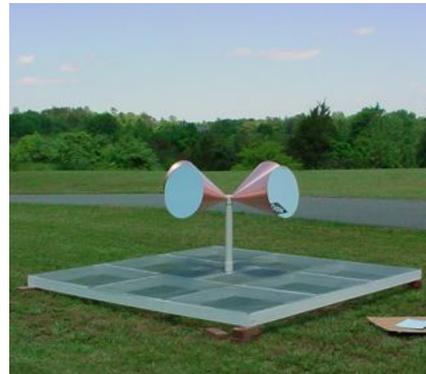
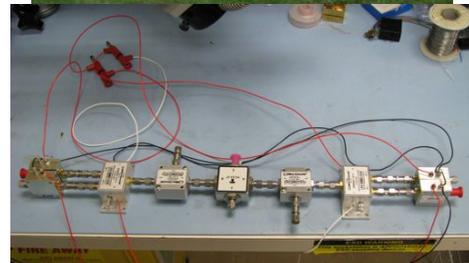

**Figure 4. (*Left*) Artist's concept of the DARE spacecraft. (*Right, top*) Prototype DARE antenna undergoing tests at NRAO. (*Right, bottom*) Prototype DARE receiver undergoing tests at JPL.**

DARE orbits the Moon for a mission lifetime of 3 years and takes science data only while above the lunar farside. The science instrument is composed of a low frequency radiometer, including electrically short, tapered, bi-conical dipole antennas, a receiver, and a digital spectrometer. The radiometer design is informed by a long heritage of ground- and space-based radio systems, and all low-level components are at a high technology readiness level (TRL).

Work on DARE proceeds on two fronts. First, a prototype system is being developed in order to increase the radiometer's TRL (Figure 4). The various sub-systems have been constructed and are in the process of system integration. Following system integration, the system will be deployed to a domestic location for initial check-out (most likely the Green Bank, WV, site of the National Radio Astronomy Observatory). Following check-out, the system will then be deployed to the Murchison Radio Observatory in Western Australia, one of the candidate sites for the international Square Kilometre Array and which is recognized as one of the most radio quiet locations on the planet.

Concurrently, the analysis techniques proposed for use with DARE are being validated using higher frequency, ground-based data. The smooth frequency response of the antennas and the differential spectral calibration approach using a Markov Chain Monte Carlo technique will be applied to detect the weak cosmic 21 cm signal in the presence of the intense solar system and Galactic foreground emissions (Harker et al. 2012).

### ii) Science Antennas for a Future Lunar Radio Telescope Array

Full exploitation of the cosmological and astrophysical information contained in the fluctuations of the highly redshifted 21 cm signal will require substantially more collecting area that can be acquired with one (or a few) orbiting spacecraft. The Lunar Radio Telescope Array (LRTA) is a concept for a telescope located in the RFI shielded zone on the far side of the Moon with a prime science mission of extracting cosmological and astrophysical measurements from power spectra of the highly redshifted 21 cm signal. While there are a number of on-going efforts to detect and study the 21 cm signals from the Hot Bubble Dominated Epoch and the Epoch of Reionization ($z < 15$), a combination of radio frequency interference (RFI) and ionospheric effects will make probing to higher redshifts increasingly more difficult.

In many aspects, the design of the LTRA will be influenced by these on-going ground-based efforts, most notably in the system architecture of the telescope and the processing of the signals. Further, there are a number of other technologies that are being developed, in many different contexts that are likely to be applicable to the LTRA. Potential examples include the development of higher-energy density batteries for power storage and increased rover autonomy for telescope array deployment. One important technology unlikely to be developed in other contexts, however, is the design for the science antennas that collect the radio radiation. Moreover, the experience of the ground-based telescope arrays is of little use because those science antennas have not been developed with the mass and volume constraints of a space mission.

LUNAR team members are engaged in technology development specifically to prove new antenna designs capable of being deployed in significant numbers on the lunar surface (Figure 5). Two primary technologies are being considered.

**Polyimide film dipoles**: Polyimide film is a flexible substance with a substantial heritage in space flight applications. In this concept, a conducting substance is deposited on the polyimide film to form the antenna. The film would then be rolled, for storage in a small volume during transport. Once on the Moon, the polyimide film rolls would be unrolled to deploy the antennas. Preliminary tests conducted on proof-of-concept antennas have demonstrated that they have the expected electromagnetic properties and that they are likely to survive the harsh lunar environment (Lazio et al. 2012).

**Magnetic helices**: In this concept, high gain antennas would be formed from helical structures. Constructed from memory metals, the antennas could be packaged in small "pizza boxes" for transport. Once on the Moon, the memory metal material would cause the antennas to assume the desired shape. Preliminary tests conducted on a proof-of-concept antenna have demonstrated that it has the expected electromagnetic properties.

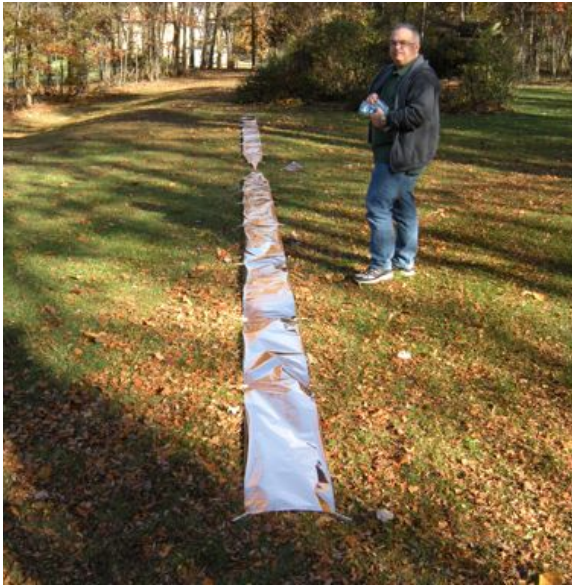 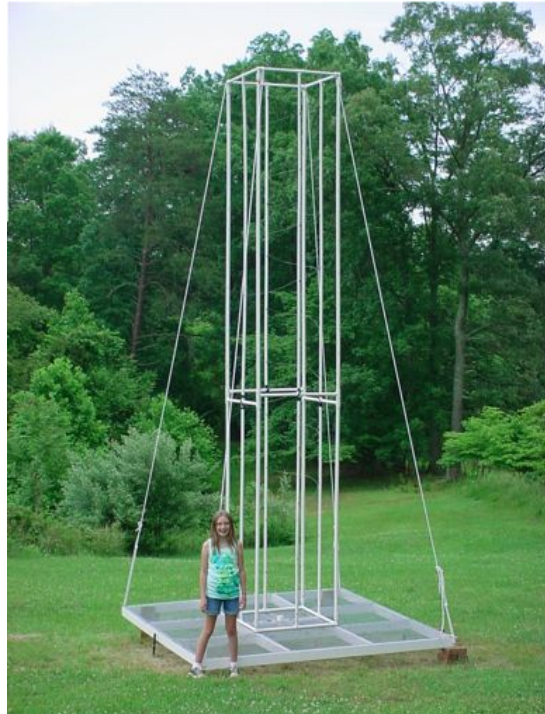

**Figure 5.** (*Left*) Polyimide film-based dipole antenna capable of being transported in a roll and deployed by unrolling on the lunar surface, being developed by NRL, NASA/GSFC, and JPL. (*Right*) Magnetic helix antenna capable of being constructed from memory metals and self deployed on the lunar surface, being developed by the NRAO. The antenna is wrapped around a support structure, which is required because of the high terrestrial gravity, but the antenna itself is not visible.

## 2) Understanding New Worlds

### a) Recommendations from *New Worlds, New Horizons in Astronomy & Astrophysics*

The *New Worlds, New Horizons in Astronomy and Astrophysics* Decadal Survey (NWNH) identified "New Worlds" as one of the three science objectives guiding the science program for the decade between 2010 and 2020. In the science program articulated in NWNH (Chapter 2), the **identification and characterization of nearby habitable exoplanets** was identified as a science frontier discovery area that could provide the opportunity for "transformational comprehension, i.e., discovery"; **"How do circumstellar disks evolve and form planetary systems?"** was identified as a science frontier question in the Origins theme; and **"How diverse are planetary systems and can we identify the telltale signs of life on an exoplanet?"** was identified as a science frontier question in the Understanding the Cosmic Order theme.

### b) Lunar Formation and Extrasolar Planetary Systems

Studies of giant planets' interaction with a protoplanetary gas disk show that their orbits migrate radially, typically achieving a compact configuration in which the pairs of neighbor planets are locked in the mean motion resonances (Kley 2000). The resonant planetary systems emerging from protoplanetary disks can become dynamically unstable after the gas disappears, leading to a phase when planets scatter off of each other. This model can explain the observed resonant exoplanets, commonly large exoplanet eccentricities (Weidenschilling & Marzari 1996), spectral features from extrasolar planetary systems (Lisse et al. 2011), and microlensing data that show evidence for a large number of planets that are free-floating in interstellar space (Sumi et al. 2011).

The solar system, with the widely spaced and nearly circular orbits of the giant planets, bears little resemblance to the bulk of known exoplanets. Yet, if our understanding of physics of planet-gas disk interaction is correct, it is likely that the young solar system followed the evolutionary path outlined above. Jupiter and Saturn, for example, were most likely trapped in the 3:2 resonance (Masset & Snellgrove 2001), defined as $P_S/P_J = 1.5$, where $P_J$ and $P_S$ are the orbital periods of Jupiter and Saturn. (This ratio is 2.49 today.)

To stretch to the present, more relaxed state, the outer solar system most likely underwent a violent phase when planets scattered off of each other and acquired eccentric orbits (Thommes et al. 1999; Tsiganis et al. 2005). The system was subsequently stabilized by damping the excess orbital energy into the transplanetary disk, whose remains survived to this time in the Kuiper belt. Finally, as evidenced by dynamical structures observed in the present Kuiper belt, planets radially migrated to their current orbits by scattering planetesimals (Malhotra et al. 1995; Levison et al. 2008).

Based on studies of the lunar cratering record, it is often assumed that the instability occurred at the time of the Late Heavy Bombardment (LHB) of the Moon, approximately 0.7 Gyr after the formation of the solar system, when the lunar basins with known ages formed (Hartmann et al. 2000). An LHB occurring in the solar system at about this time is consistent with emerging evidence from extrasolar planetary systems that show spectral features consistent with significant collisional evolution (Lisse et al. 2011). If so, the solar system's giant planets were required to remain on their initial resonant orbits for up to 700 Myr, a significant constraint on models of the early solar system.

Interestingly, the models starting with a resonant system of four giant planets have a low success rate in matching the present orbits of giant planets and various other constraints (e.g., survival of the terrestrial planets, Nesvorny et al. 2011). The dynamical evolution is typically too violent, if Jupiter and Saturn start in the 3:2 resonance, and leads to final systems with fewer than four planets.

Instead, some of the statistically best results were obtained when assuming that the solar system initially had five giant planets and one ice giant, with the mass comparable to that of Uranus and Neptune, was ejected to interstellar space by Jupiter (Figure 6). This possibility appears to be conceivable in view of the recent discovery of a large number of free-floating planets in interstellar space, which indicates that planet ejection should be common (Sumi et al. 2011).

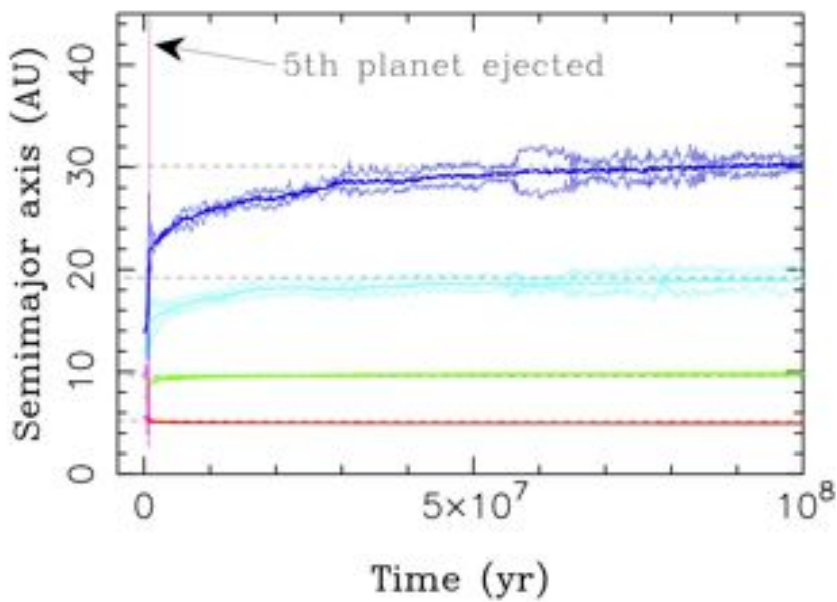

Figure 6. Orbit histories of giant planets in a simulation of our solar system with five initial planets. The five planets started in the (3:2, 3:2, 4:3, 5:4) resonances. After a series of encounters with Jupiter, the inner ice giant was ejected from the solar system at $8.2 \times 10^5$ yr (purple path). The remaining planets were stabilized by the planetesimal disk and migrated to orbits that very closely match the current orbits of the outer planets (dashed lines). (From Nesvorny et al. 2011)

### c) Magnetospheric Emissions from Extrasolar Planets

The magnetic polar regions of the Earth and the solar system giant planets host intense electron cyclotron masers generated by interactions between solar wind-powered currents and planetary magnetospheric fields. Empirical relations for solar system planets suggest that extrasolar planetary radio emission may be detectable (Farrell, Desch, & Zarka 1999), and the emission frequencies of some planets may be in the range relevant to Cosmic Dawn studies (e.g., Lazio et al. 2010).

The dynamo currents generating a planet's magnetic field arise from differential rotation, convection, compositional dynamics, or a combination of these in the planet's interior.

Consequently, knowledge of the planetary magnetic field places constraints on a variety of planetary properties, some of which will be difficult to determine by other means.

- **Planetary Interiors**: For the solar system planets, the composition of the conducting fluid ranges from liquid iron in the Earth's core to metallic hydrogen in Jupiter and Saturn to perhaps a salty ocean in Uranus and Neptune. Radio detection of an extrasolar planet, combined with an estimate of the planet's mass and radius, would indicate the planet's internal composition, by analogy to the solar system planets.
- **Planetary rotation**: The rotation of a planet imposes a periodic modulation on the radio emission, as the emission is preferentially beamed in the direction of the local magnetic field and will change if the magnetic and spin axes of the planet are not aligned. For all of the gas giant planets in the solar system, this modulation defines their rotation periods.
- **Planetary Satellites**: In addition to being modulated by its rotation, Jupiter's radio emission is affected by the presence of its satellite Io, and more weakly by Callisto and Ganymede. Modulations of planetary radio emission may thus reveal the presence of a satellite.
- **Atmospheric retention**: A common and simple means of estimating whether a planet can retain its atmosphere is to compare the thermal velocity of atmospheric molecules with the planet's escape velocity. If the thermal velocity is a substantial fraction of the escape velocity, the planet will lose its atmosphere. For a planet immersed in a stellar wind, non-thermal atmospheric loss mechanisms can be important (Shizgal & Arkos 1996), as the typical stellar wind particle has a supra-thermal velocity. If directly exposed to a stellar wind, a planet's atmosphere can erode more quickly. Based on Mars Global Surveyor observations, this erosion process is thought to have been important for Mars' atmosphere and oceans (Lundin et al. 2001; Crider et al. 2005).
- **Habitability**: A magnetic field may determine the habitability of a planet by deflecting cosmic rays or stellar wind particles, (e.g., Griessmeier et al. 2005). In addition to its effect on the atmosphere, if the cosmic ray flux at the surface of an otherwise habitable planet is too large, it could cause cellular damage or frustrate the origin of life altogether.

Absent a direct radio detection of magnetically generated emission, evidence for extrasolar planetary magnetic fields is suggestive, but ambiguous, ranging from efforts to detect aurorally generated UV emission from "hot Jupiters" (France et al. 2010) to efforts to explain the apparently inflated radii of "hot Jupiters" by Ohmic dissipation (e.g., Batygin & Stevenson 2010) to stars showing apparently magnetically-generated star-planet interactions (Shkolnik et al. 2008). However, the range of magnetic field strengths varies immensely, from 10% that of Jupiter to several times that of Jupiter, sometimes for the same object!

# 3) Testing the Physics of the Universe: Gravitation

**Project Leaders:**  Dr. Douglas Currie (University of Maryland, College Park)
Dr. Stephen Merkowitz (NASA/Goddard Space Flight Center)
Dr. Thomas Murphy (University of California, San Diego)

### a) Recommendations from *New Worlds, New Horizons in Astronomy & Astrophysics*

The *New Worlds, New Horizons in Astronomy & Astrophysics* Decadal Survey (NWNH) identified "Physics of the Universe" as one of the three science objectives guiding the science program for the decade between 2010 and 2020. In the science program articulated in NWNH (Chapter 2), "**Why is the Universe accelerating?**" and **"What is dark matter?"** were identified as science frontier questions in the Frontiers of Knowledge theme.

The NWNH Particle Astrophysics and Gravitation (PAG) Program Prioritization Panel expanded upon this science objective by noting the many constraints on general relativity already provided by lunar laser ranging (described in more detail below) and describing lunar laser ranging as a "cost effective" means of "testing general relativity as accurately as possible." Further, they explicitly described lunar laser ranging as an important component of a system for testing fundamental physics, particularly "if conducted as a low-cost robotic mission or an add-on to a manned mission to the Moon."

Similar recommendations appear in the report *The Scientific Context for Exploration of the Moon*, which recommended that retroreflectors for lunar laser ranging should be one of the instrument packages on future landers.

### b) Approach of the LUNAR Team

The Lunar Laser Ranging (LLR) component of the LUNAR team has taken a two-fold approach toward testing theories of gravity. The first is to continue a long-standing program of precise measurements of the Earth-Moon distance via the technique of LLR. Because the sensitivity of the LLR technique increases with the duration of the observing program, LLR constraints on deviations from general relativity (GR) become increasingly austere. Second, the LUNAR team is leading efforts to develop a next-generation retrore-flector package (e.g., the Lunar Laser Ranging Retroreflector Array for the 21st Century or LLRRA-21) so that future missions to the Moon can emplace new retroreflectors. Doing so, could increase both the latitude and longitude coverage of retroreflectors and thereby improve significantly the precision of the constraints.

### c) Background on Lunar Laser Ranging (LLR) Technique

Forty-two years ago, the first retroreflector was placed on the surface of the Moon by the *Apollo* 11 astronauts. Principal Investigator on the development of those retroreflector arrays was LUNAR team member D. Currie.

These retroreflectors provided fixed-point targets on the lunar surface and reflected the laser signal directly back to the telescope that was being used as both the laser transmitter and return signal receiver. They were designed to operate in both the lunar night and the lunar day with a long lifetime. These retroreflectors are still producing excellent data for monitoring the Earth-Moon separation, and the LLR program, in large part by virtue of its long duration, has been one of the major contributors in the study and experimental exploration of gravitational physics.

Over the past two decades, new discoveries have been made that are extremely difficult to explain within the structure of GR. In particular these consist of

- **The Inability to "Quantize" GR**: As a classical theory, GR and quantum mechanics are fundamentally inconsistent. There must be a breakdown at some level of accuracy in GR or a problem with quantum mechanics.
- **The Existence of Dark Energy:** The 2011 Nobel Prize recognized the discovery of an accelerating expansion of the Universe. While this expansion can be accommodated within GR by a cosmological constant, or more generally by "dark energy," the current (small) value of the cosmological constant seems to have little justification.
- **The Existence of Dark Matter:** The inferred amount of dark matter vastly exceeds that provided by known species of neutrinos. There might be other species of sub-atomic particles or the assumption that GR applies equally on all length scales may not be correct.

Each of these problems has led to the creation of a plethora of hypotheses to attempt to explain the issues, a number of which already have already been discarded because they are not consistent with the current measurements provided by the LLR program (and other measures of GR). Continuing the LLR program is essential to provide increasingly stringent constraints that current or future hypotheses must meet.

### i) History of the *Apollo* Arrays

The physics of the candidate retroreflectors (Figure 7; Chang et al. 1972) and then the development of the *Apollo* 11 array (Alley et al. 1970) was addressed by a national team. The team was centered at the University of Maryland, and it developed the retroreflector array as an element of the Early Lunar Science Experiment Program (ELSEP). The retroreflector consisted of a panel of 100 cube corner reflectors (CCRs). After deployment, ranging to the *Apollo* 11 retroreflector array was then successfully accomplished at the Lick Observatory and in a continuing program at the McDonald Observatory of the University of Texas (Bender et al. 1973). At the Jet Propulsion Laboratory (JPL), the ranging data was used to greatly improve the lunar ephemeris (from kilometers to meters) and to start to extract the various aspects of gravitational physics from the data (Williams et al. 2004).

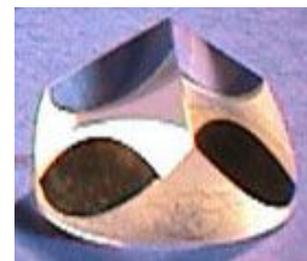

Figure 7. Solid cube corner reflector (CCR)

Based on this early success, additional retroreflector arrays were deployed. Figure 8

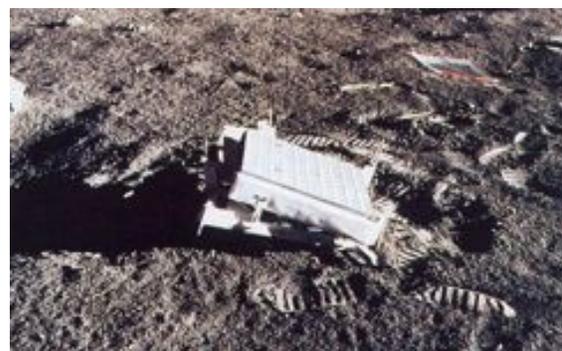

Figure 8. *Apollo* 14 LLR deployed on the lunar surface.

shows the second *Apollo* array, emplaced during the *Apollo* 14 mission. Finally, an upgraded array with 300 CCRs was developed and deployed during *Apollo* 15. A minimum of three arrays is necessary to track the rotation (i.e., librations) of the Moon and determine the center of mass (CoM) of the Moon and a variety of other features of the lunar interior with high accuracy. In turn, the knowledge of the position of the Moon's CoM is used for the gravitational physics tests.

### ii) History of Lunar Laser Ranging Program

Over the years following the deployment of the *Apollo* arrays, and two similar arrays by the Soviet space program (Figure 9), regular ranging, typically three times a day, was conducted at the McDonald Observatory, supplemented by observations at various other observatories (Figure 10). The latter is important as observations from multiple stations on the Earth define various parameters related to the Earth's CoM as it relates to the determination of the Earth-Moon distance, defined as the distance between their CoMs. At JPL, the continuing data were used to improve the ephemeris of the Moon. Over the decades, the initial accuracy of the ephemeris as compared to the ranging improved from approximately 300 mm to approximately 20 mm. As will be discussed later, this limiting value is due the lunar librations tilting the arrays so the return pulse is spread in time over the equivalent of 20 mm, even if the up-going laser pulse had a spread equivalent to only 1 mm. Thus, neither the ephemeris nor the ranging stations had reason to improve further.

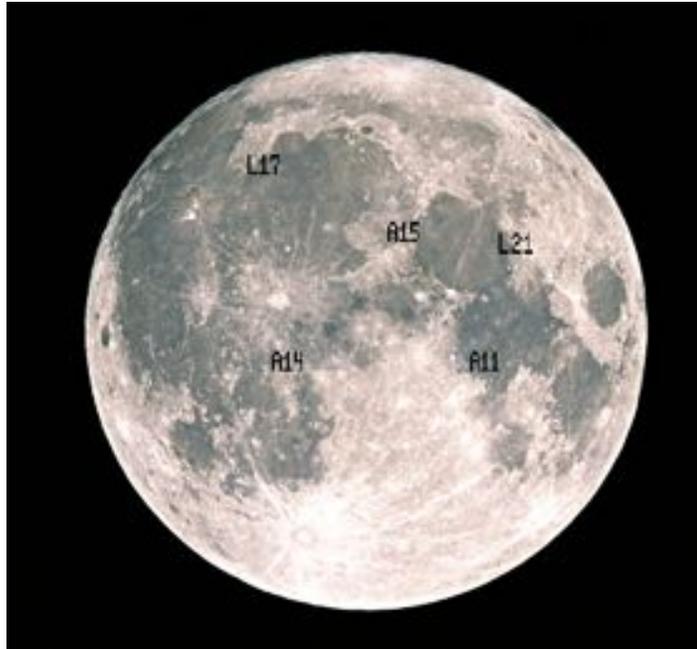

**Figure 9. Placement of three Apollo LLR and the two Soviet LLR (L1 & L2).**

At JPL, the LLR program continued to constrain the physics of gravitation by a detailed comparison between the observations of the distance to the Moon and the model of the range. This is accomplished by modifying the GR parameters (and other parameters) to minimize the residuals (observations – model). Over the decades, the result has been some of the best values for evaluating various theories that have been proposed to replace GR.

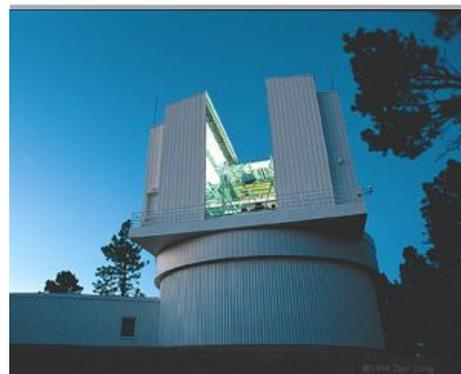

**Figure 10. APOLLO lunar laser ranging station, currently the premier ranging station.**

### iii) Range Residuals

The original objective of the LLR program was measurements at the 300 mm level (Figure 11). As ground stations have improved, we are currently limited to 20 mm by lunar librations. Thus over the past two decades, the accuracy to which we are able to improve the modeling has been limited to about the 20 mm level.

### d) Scientific Results of the Apollo-based Lunar Laser Ranging Program

There are a wide range of science results that can be extracted from the detailed lunar ephemeris that has been obtained in forty years of LLR and the data processing at the JPL. Here we shall concentrate on a few of the more interesting astrophysical results.

#### A) Inertial Properties of Gravitational Energy

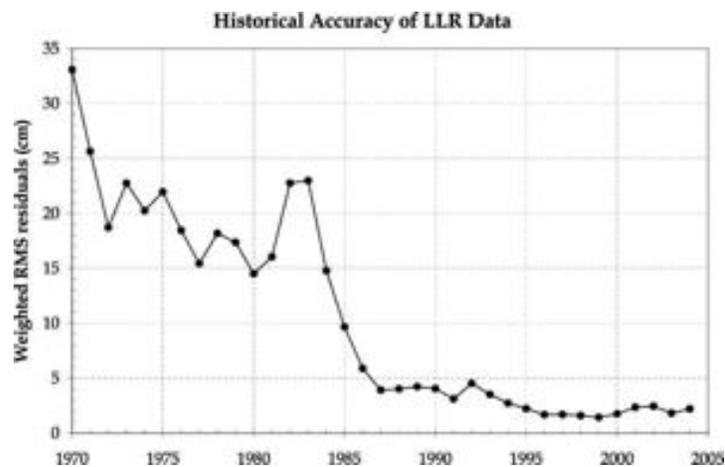

Figure 11. (Range Observation – Model) residuals showing the improvement in ground stations over the past four decades.

One formulation of the Equivalence Principle is that all bodies fall at the same rate, regardless of their composition. GR states that gravitational binding energy has mass and thus will contribute to the inertial properties of a body. By comparison with the "rate of falling toward the Sun" of the Earth and Moon, the LLR program has demonstrated that gravitational binding energy has inertial properties like baryonic matter, to a precision of a few parts in $10^{-4}$.

#### B) Change of the Gravitational Constant with Time

One class of proposed theories to replace GR involve temporal and spatial variation in the constants of nature, particularly the gravitational constant G. With the relatively long observational record (i.e., more than 40 years), the LLR program is particularly adapted to evaluate the change in G with time. At present, limits on the variation of G are $(4 \pm 6) \times 10^{-13}$ yr$^{-1}$, less than 1% over the entire life of the Universe.

#### C) Change of the Gravitational Constant with Space

The other constant of nature tested by LLR is any spatial change in the gravitational constant on the scale for a few hundred megameters.

The regular observations are conducted with an accuracy of each measurement of the distance to the moon of a few parts in $10^{-12}$. More precisely, other parameters are constrained as follows

- Limit on a violation of the weak equivalence principle (WEP) difference between the ratio of the gravitational binding energy and inertial masses for the Earth and Moon is $(-1.0 \pm 1.4) \times 10^{-13}$ (Williams et al. 2005);

- The post-Newtonian parameter (PPN) β is constrained to be $\beta - 1 = (1.2 \pm 1.1) \times 10^{-4}$ (Williams et al. 2005);
- The strong equivalence principle (SEP) violation parameter η, which itself is a linear combination of PPN parameters, is constrained to be $\eta = 4\beta - \gamma - 3 = (4.4 \pm 4.5) \times 10^{-4}$ (Williams et al. 2005);
- The deviation of geodetic precession from its GR value is constrained to be $K_{gp} = (1.9 \pm 6.4) \times 10^{-3}$ (Williams et al. 2004);
- Any temporal variation in the gravitational constant is constrained to be $(4 \pm 6) \times 10^{-13}$ yr$^{-1}$ (Williams et al. 2004);
- Limits on deviations of the inverse square law on scales of $10^8$ m are constrained to better than $10^{-10}$ (Murphy 2009); and
- Limits on gravitomagnetism (e.g., frame dragging) to better than 0.1% (Murphy 2009).

### e) Current Limitations of the *Apollo* Arrays

The three retroreflector arrays that were deployed during the *Apollo* 11, 14, and 15 missions have continued to operate for the past forty two years. During this time, the accuracy of the ground stations has improved by more than two orders of magnitude. Thus, the properties of the *Apollo* arrays are now one of the limiting effects in our ability to extract more science from LLR. In particular, each of the *Apollo* arrays is a panel of 100 or 300 cube corner reflectors. As the Moon librates, the panel is tilted so the nearest corner cube is closer to the Earth with respect to the other corner by as much as 25 mm, meaning that a single photoelectron return has an uncertainty that is many millimeters.

In addition, the reduced return from the *Apollo* 11 and 14 arrays means that few ground stations can observe all three arrays in a short time, a procedure that is critical in obtaining data the optimized the data processing.

Details of the original retroreflectors now limit the accuracy for continuing to improve the accuracy in the study of the structure of GR in comparison with the other proposed theories of gravitation. However, a continuing evaluation of the structure of GR is crucial in addressing the fundamental inconsistency of the two theories that are the basis of our understanding of nature of the physical world, that is, GR and Quantum Mechanics.

### f) Next Generation Lunar Laser Retroreflector

A next generation retroreflector will be essential to improving the precision obtained from LLR, both for gravitational physics and for understanding the lunar interior. Key objectives for next generation retroreflectors include

#### A. Improve the Ranging Accuracy Supported by the Lunar Emplacements

The objective is to provide a lunar emplacement that will support a ranging accuracy well beyond the current single shot accuracy of about 20 mm. The new retroreflectors will support much higher accuracy (< 100 μm) to support the improvement in the LLR ground stations over the next decades. Since the *Apollo* retroreflector arrays were deployed, the ground stations have improved their accuracy by more than two orders of magnitude. This will lead to an improvement in the scientific results in the testing of GR to more than two orders of magnitude.

## B. Enable Access to the High Accuracy Ranging for Multiple Ground Stations

The new retroreflectors will have signals that can be accessed by an improved number of LLR ground stations, especially those that are dedicated to regular daily ranging. At present, there are a number of stations can access *Apollo* 15 arrays but not the *Apollo* 11 and 14 arrays. In particular, this will enable the role of smaller telescopes that can be dedicated to LLR. The new retroreflectors will allow the additional regular ranging stations to have access to multiple arrays so that
a. There will be a greater rate of ranges that can enter the analysis program to improve the achievable accuracy,
b. We can establish the parameters of related to the properties of the Earth and the precise station locations,
c. By having multiple ground stations, we can achieve an improved understanding of any systematic errors affecting a given ground station.

A leading candidate for the next-generation LLR is the "Lunar Laser Ranging Retroreflector for the 21st Century" (a.k.a. LLRRA-21, Figure 12) being developed at the University of Maryland. The magnitude of the improved accuracy depends upon the method of deployment, i.e., on a lander, on the regolith or an anchored deployment, drilling down about one meter and locking to the regolith at this point. This approach has a very high heritage that bodes well for a highly successful program. In particular, both the scientific heritage, illustrated by the above results from the operating *Apollo* arrays, and a technology heritage, illustrated by the successful operation of the *Apollo* arrays on the lunar surface for more than four decades, and by the thousands of solid cube corner retroreflectors in low and intermediate altitude in Earth orbit.

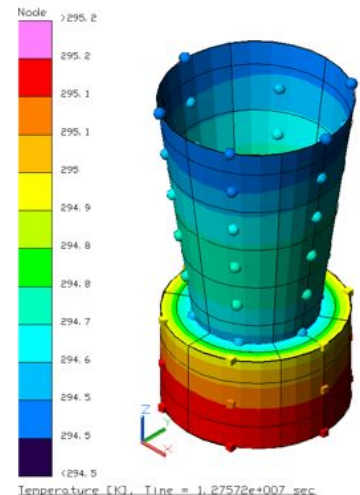

**Figure 12. Model of LLRRA-21, as operating in the thermal modeling of lunar performance.**

## Acknowledgements

The LUNAR consortium and the CLOE consortium are both funded by the NASA Lunar Science Institute. Part of this research was carried out at the Jet Propulsion Laboratory, California Institute of Technology, under a contract with the National Aeronautics and Space Administration.